\documentclass[aps,prb,twocolumn,amsmath, amssymb]{revtex4-1}

\usepackage{float}
\usepackage{graphicx}
\usepackage{braket}
\usepackage{ulem}   
\usepackage{dsfont}
\usepackage{amsthm,amsmath,amsfonts,amssymb,verbatim,color}
\usepackage{bbold}
\usepackage{mathtools}
\usepackage{graphicx}
\usepackage[T1]{fontenc}
\usepackage[colorlinks=true,citecolor=blue,linkcolor=blue,urlcolor=blue]{hyperref}
\newcommand{\bs}[1]{{\boldsymbol{#1}}}

\newcommand{\tr}{\mathop{\mathrm{Tr}}}

\newcommand{\im}{\mathop{\mathrm{Im}}}

\begin{document}
\preprint{}
\title{Topological Transitions in
Orbital-Symmetry-Controlled Chemical Reactions}

\author{Ziren Xie}
\affiliation{Department of Chemistry, The Pennsylvania State University,  University Park 16802, USA}
\author{Amir Mirzanejad}
\affiliation{Department of Chemistry, The Pennsylvania State University,  University Park 16802, USA}
\author{Lukas Muechler}
\affiliation{Department of Chemistry, The Pennsylvania State University,  University Park 16802, USA}
\affiliation{Department of Physics, Pennsylvania State University, University Park, Pennsylvania 16802, USA}
\email{lfm5572@psu.edu}

\begin{abstract}	
Topological band theory has transformed our understanding of crystalline materials by classifying the connectivity and crossings of electronic energy levels. Extending these concepts to molecular systems has therefore attracted significant interest.  Reactions governed by orbital symmetry conservation are ideal candidates, as they classify pathways as symmetry-allowed or symmetry-forbidden depending on whether molecular orbitals cross along the reaction coordinate. However, the presence of strong electronic correlations in these reactions invalidate the framework underlying topological band theory, preventing direct generalization. Here, we introduce a formalism in terms of Green's functions to classify orbital symmetry controlled reactions even in the presence of strong electronic correlations. Focusing on prototypical 4$\pi$ electrocyclizations, we show that symmetry-forbidden pathways are characterized by crossings of Green's function zeros, in stark contrast to the crossings of poles as predicted by molecular-orbital theory.  We introduce a topological invariant that identifies these symmetry protected crossings of both poles and zeros along a reaction coordinate  and outline generalizations of our approach to reactions without any conserved spatial symmetries along the reaction path. Our work lays the groundwork for systematic application of modern topological methods to chemical reactions and can be extended to reactions involving different spin states or excited states.
\end{abstract}
\date\today
\maketitle

%

\section{Introduction}
Recent developments in topological band theory have systematically classified all possible band structures in crystalline materials~\cite{Chiu2016-nm,Bradlyn2017-zu,watanabe2017structure,Po2017-gw,Kruthoff2017-sf,Vergniory2019-sy,Zhang2019-is,Vergniory2022-ov}.
The classification is achieved via topological invariants, non-local order parameters calculated from the electronic wavefunctions, first made rigorous by Berry~\cite{Berry1984-ag}. Invariants predict the existence of topological surface or interface states that can be measured in experiment and predicted with high accuracy from first principles calculations~\cite{Bansil2016-wb,Manna2018-fn,Xiao2021-mh,Lv2021-pb,Kumar2021-az}.
Further, they facilitate the classification of crossings or avoided crossings of energy levels, which can occur at isolated points, lines or surfaces in reciprocal space.

A classification through invariants is inherently resistant to small changes and perturbations of the system. A quantum phase transition is required to change them, requiring significant changes to system parameters. 
Materials that have the same topological invariant belong to the same universality class, meaning that they possess the same topological states even if they are chemically distinct. 
Recent work has highlighted how spatial symmetries enrich this classification~\cite{Chiu2014-lz,Bradlyn2016-sg,Chiu2016-nm,Kruthoff2017-sf,Po2017-gw,Song2018-ac,Po2020-ra,Wieder2021-fc}
Computed from the symmetry properties of wavefunctions, symmetry indicators allow for efficient and inexpensive calculation of topological invariants and enable high-throughput exploration of material databases~\cite{Zhang2019-is,Tang2019-wj,Vergniory2019-sy,Vergniory2022-ov}.
\begin{figure*}[t]
    \includegraphics[width=0.99\textwidth]{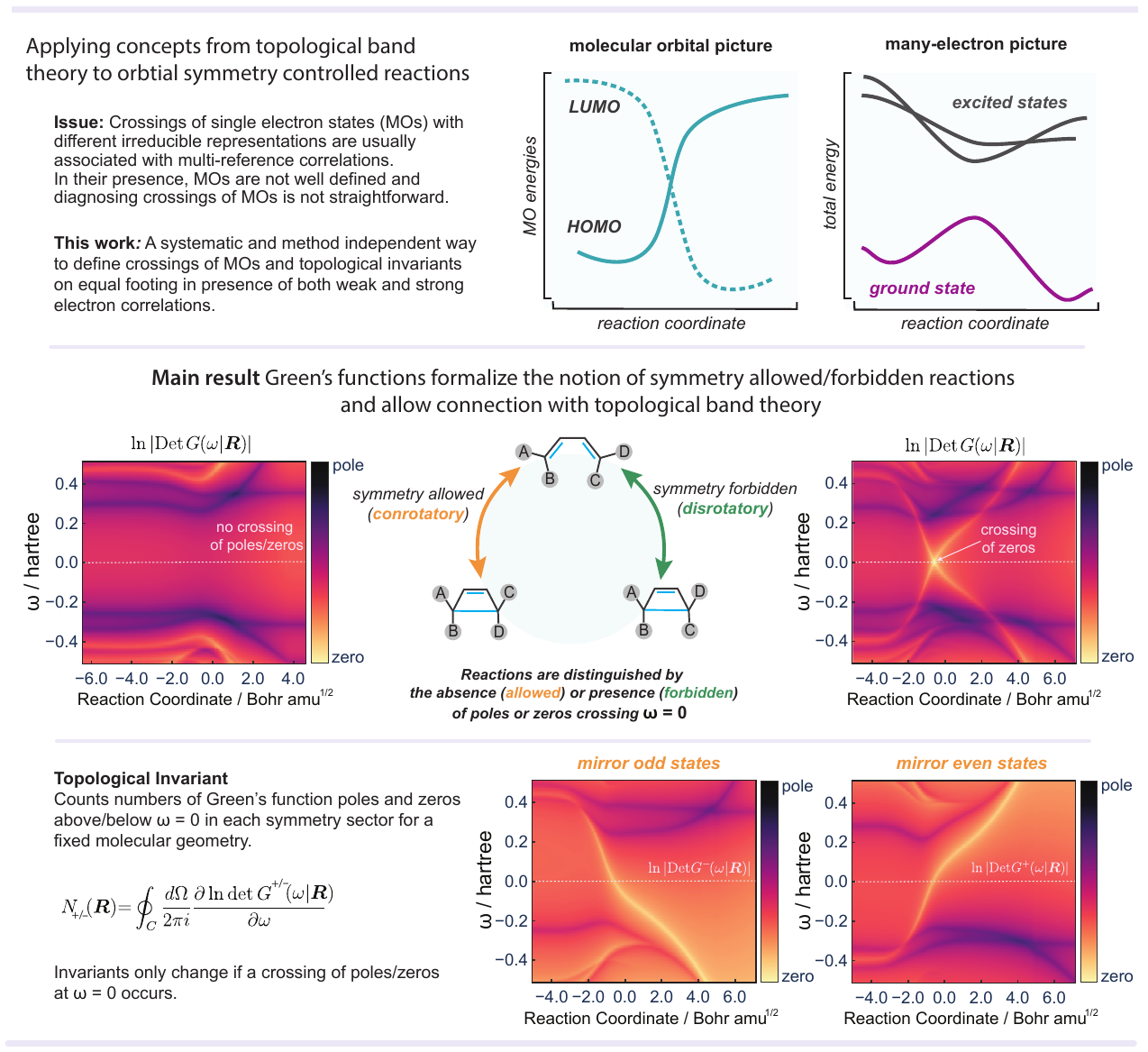}
    \caption{Summary of main results of this paper}
    \label{fig:summaryFigure}
\end{figure*}
Crossings as well as avoided crossings of energy levels are of fundamental importance to our understanding of chemical reactions. Topological concepts such as Berry phases and extensions have been applied to molecules\cite{Faure2000-fc,Zhilinskii2001-sw,Faure2002-zf,Muechler2014-nq,Dhont2017-aj,Wu2020-xj,Wu2021-qa,Bian2022-vt,Culpitt2021-cd,Culpitt2024-jf,Morreale2024-qv}, and are vital in understanding nuclear dynamics close to conical intersections, which have recently been revisited in the framework of exact factorization methods~\cite{ribeiro2018vibronic,Ryabinkin2017-jn,Ibele2023-ij,Requist2016-ib,Requist2017-yk,Valahu2023-pn}.

The interplay of energy level crossings and symmetries on the other hand is well established in symmetry controlled reactions such as those described by the Woodward-Hoffmann rules (WHR) and generalizations thereof~\cite{Woodward1969-ds,Waschewsky1994-er,Butler1998-ek,Muechler2020-de,Zhou2024-ji,Ong2009-sn,Kochhar2010-nj,Brown2021-if}. Reactions are classified as \textit{allowed} when no crossing of molecular orbitals (MOs) occur along the reaction coordinate. Reactions with a crossing between occupied and unoccupied MOs are classified as \textit{forbidden}.
This apparent connection between energy level crossings and symmetries suggests symmetry controlled reactions as an ideal platform to study the application of modern topological methods in reaction chemistry.

However, the application of topological concepts to these reactions is not straightforward. A principal limitation is that topological band theory is typically formulated within a single-electron or single-reference framework.
While molecules close to their equilibrium geometry can often be described by single-reference methods, taking into account electron-electron correlations explicitly is especially important for chemical reactions when bonds are broken and formed~\cite{Boguslawski2013-ga,Krylov2017-vh,Park2020-sz,Zhou2024-ji}.
In particular, when HOMO and LUMO are degenerate or near-degenerate, as in symmetry forbidden reactions, electronic interactions become significant, and single-electron approximations break down. 
The limitations of the single-electron picture in the presence of strong static correlations present a challenge for a topological analysis. For example, in symmetry forbidden reactions, MO theory predicts a crossing of occupied and unoccupied MOs with different irreducible representations (Fig.~\ref{fig:summaryFigure}). 
When electron correlation is explicitly included using multi-reference methods, the MO crossing is no longer well-defined~\cite{Longuet-Higgins1965-xc,Goddard1972-qx,Ross1979-je}. Instead, there is an avoided crossing between the low-lying many-electron singlet states ($S_0,S_1,S_2$). While this accurately describes the energetics, forbidden reactions are no longer distinguished from allowed reactions, as both feature a gapped ground-state along the reaction coordinate.

To address this limitation and facilitate the application of topological tools to a wider range of chemical systems, we introduce a theoretical framework based on Green's functions developed recently for strongly correlated crystalline materials. This method is capable of describing both weakly and strongly correlated electronic systems on a consistent basis, and allows to diagnose the crossings of molecular orbitals even when strong multi-reference correlations are present.

\section{Results}
\subsection{Green's functions and Multi-Reference Correlations}
Green's functions are central objects in electronic structure theory
~\cite{cederbaum1977theoretical,schirmer1983new,von1984computational,rusakov2016self,phillips2015fractional,jacquemin2017bethe,van2015gw,golze2019gw,van2013gw,Laughon2022-yq,Phillips2014-do}, the theory of strongly correlated electrons~\cite{Luttinger1960-xr,Georges1996-gj,Onida2002-vn,zgid2011dynamical,lin2011dynamical,Shee2019-ub,Mahan2010-dr}
and topological condensed matter physics~\cite{gurarie2011single,Slager2015-zr,Seki2017-hq,lessnich2021elementary,iraola2021towards,Misawa2022-hk,soldini2023interacting,Peralta-Gavensky2023-bk,Queiroz2024-cq}.
The Green's Function describes single particle properties of molecules, such as excitations and electronic correlations, and can be measured in experiment.
For example, the spectral function \mbox{$A(\omega|\bs{R}) = \lim_{\gamma\rightarrow0} \im \tr \left[G\left(\omega +i\gamma | \bs{R}\right)\right]$} is measured photoemission and inverse photoemission experiments.
Assuming a non-degenerate singlet ground state $\ket{S_0}$ with $N$ electrons, the Green's function in the Lehmann representation for a fixed set of nuclear coordinates $\bs{R}$ is given as
\begin{equation}
\begin{split}
    G_{ij}^{\sigma\sigma'}\left(\omega | \bs{R}\right)& =\sum_n \frac{\braket{S_0(\bs{R})| c_{i\sigma}|n(\bs{R}}\braket{n(\bs{R})| c_{j\sigma'}^{\dagger}|S_0(\bs{R})}}{\omega-\left[(E_n(\bs{R},N+1)-E_{S_0}(\bs{R},N)\right]} \\ 
        &+\sum_m \frac{\braket{S_0(\bs{R})| c_{j\sigma'}^{\dagger}|m(\bs{R})} \braket{m(\bs{R})| c_{i\sigma}|S_0(\bs{R})}}{\omega-\left[(E_{S_0}(\bs{R},N)-E_m(\bs{R},N-1)\right]}
        \label{eq:GF}
\end{split}
\end{equation}
where $i,j$ label single electron orbitals,$\sigma$ is the electronic spin and $\Omega = \omega + i \gamma$ is the frequency in the complex plane.
$E_{S_0}(\bs{R},N)$ and $\ket{S_0(\bs{R}}$ are the ground state energy and wavefunction of the system with $N$ electrons, while $\ket{n(\bs{R})},E_n(\bs{R},N+1)$ and $\ket{m(\bs{R})},E_m(\bs{R},N-1)$ are the wavefunctions and energies of the excited states of the system with $N+1$ and $N-1$ electrons respectively. We are only concerned with the spin-diagonal part $\sigma=\sigma'$, i.e. $G^{\sigma\sigma}\left(\omega | \bs{R}\right) \equiv G\left(\omega | \bs{R}\right)$ since we neglect spin-orbit coupling.

Quantum chemistry distinguishes between two limiting types of electron correlation: dynamic and static~\cite{Bulik2015-xf,jensen2017introduction,Krylov2017-vh,Park2020-sz}. Dynamic correlation arises from the correlated motion of electrons and is well described by perturbative methods. Static correlations, also known as multi-reference or strong correlations, occur when multiple electronic configurations are nearly degenerate, e.g. in systems with near-degenerate occupied and unoccupied orbitals, e.g. diradicals. In such cases, single-electron and perturbative methods fail to describe the electronic structure of the material both qualitatively and quantitatively.
In non-interacting molecules or those described by single-electron methods, the Green's function contains only poles at single-particle orbital energies: occupied orbitals correspond to poles at $\omega < 0$, and unoccupied orbitals to poles at $\omega > 0$. The spectral function is a series of delta functions at these energies.
When electron-electron interactions are fully taken into account, the self-energy $\Sigma(\omega|\bs{R}) = G_0^{-1}(\omega|\bs{R}) - G^{-1}(\omega|\bs{R})$\footnote{The self-energy depends on the definition of $G_0(\omega|\bs{R})$. Here, we calculate  $G_0(\omega|\bs{R})$ from single-particle part of the Hamiltonian obtained from CAS calculations} accounts for correlation effects. It shifts pole positions, indicative of dynamic correlations. When the self-energy diverges at a frequency, it produces a zero in the Green’s function. The appearance of these zeros close to $\omega = 0$ indicates strong static correlations. How zeros emerge and disperse as a function of system parameters have attracted considerable interest in the context of Luttinger’s theorem, particularly in the study of Mott insulators and high-temperature superconductors~\cite{Luttinger1960-xr,Dzyaloshinskii2003-wz,Stanescu2007-gs,Dave2013-dx,Kozik2015-fn,Seki2017-hq,Heath2020-ow,Fabrizio2022-qv,Wagner2023-bp,Bollmann2024-ew}. Moreover, recent work has highlighted their relevance in the theory of topological invariants for strongly correlated materials~\cite{gurarie2011single,iraola2021towards,Zhou2020-zf,Wang2022-mo,Blason2023-zp,soldini2023interacting,Mai2024-tp,Setty2024-dy}.

\subsection{First Principles Calculations}
To illustrate this concept, we turn to benzene and cyclobutadiene. In the framework of H\"uckel theory, benzene has no degeneracy between occupied and unoccupied orbitals. In contrast, cyclobutadiene in its high symmetry $D_{4h}$ geometry, possesses two degenerate, half-filled orbitals.
\begin{figure}[t]
    \includegraphics[width=0.99\columnwidth]{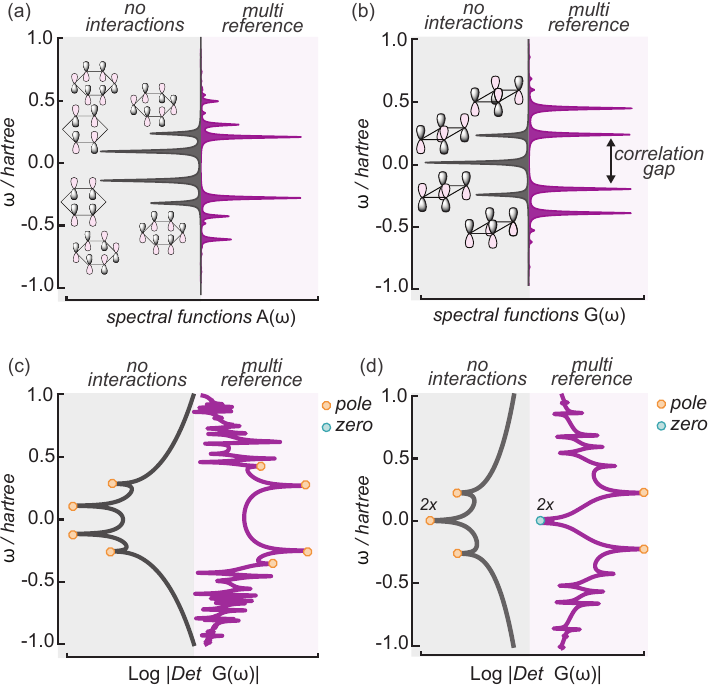}
    \caption{Spectral functions and Green’s functions plotted as $\ln|\text{Det} G\left(\Omega | \bs{R}\right)|$ of cyclobutadiene and benzene at their $D_{4h}$ and $D_{6h}$ geometries. (a) Spectral function of benzene. (b) Spectral function of cyclobutadiene. (c) Green's function of benzene. (d) Green's function of cyclobutadiene. The Green's functions were obtained from CASSCF(4,4)/def2-SVP (cyclobutadiene) and CASSCF(6,6)/def2-SVP (benzene) calculations.
    }
    \label{fig:spectralGapping}
\end{figure}
The non-interacting spectral function for benzene possesses peaks that correspond to the $\pi$ orbitals predicted by Hückel theory [Fig.~\ref{fig:spectralGapping}a], i.e. two peaks for $\omega < 0$, and two peaks for $\omega > 0$, while the Green's function possesses no zeros [Fig.~\ref{fig:spectralGapping}c].
When interactions are explicitly included the peak structure remains qualitatively unchanged for the peaks closest to $\omega = 0$. Multiple zeros emerge, but none close to $\omega = 0$. This indicates that dynamic correlations dominate in benzene.
The non-interacting spectral function of cyclobutadiene similarly features  A bonding ($\omega < 0$) and an anti-bonding orbital ($\omega > 0$), in addition to two degenerate non-bonding orbitals at $\omega = 0$ in accordance with H\"uckel theory [Fig.~\ref{fig:spectralGapping}b].
In contrast to benzene, the interacting spectral function differs qualitatively from the non-interacting. A correlation gap opens, as the two peaks at $\omega = 0$ both split into an occupied state ($\omega < 0$) and an unoccupied state ($\omega > 0$). The gap opening is accompanied by two zeros of the Green's function at $\omega = 0$ [Fig.~\ref{fig:spectralGapping}d], indicating that static-correlations dominate in cyclobutadiene.

\subsection{Green's Functions of Symmetry-Allowed and Symmetry-Forbidden Electrocyclic Reactions}
\begin{figure}[htp]
    \includegraphics[width=0.99\columnwidth]{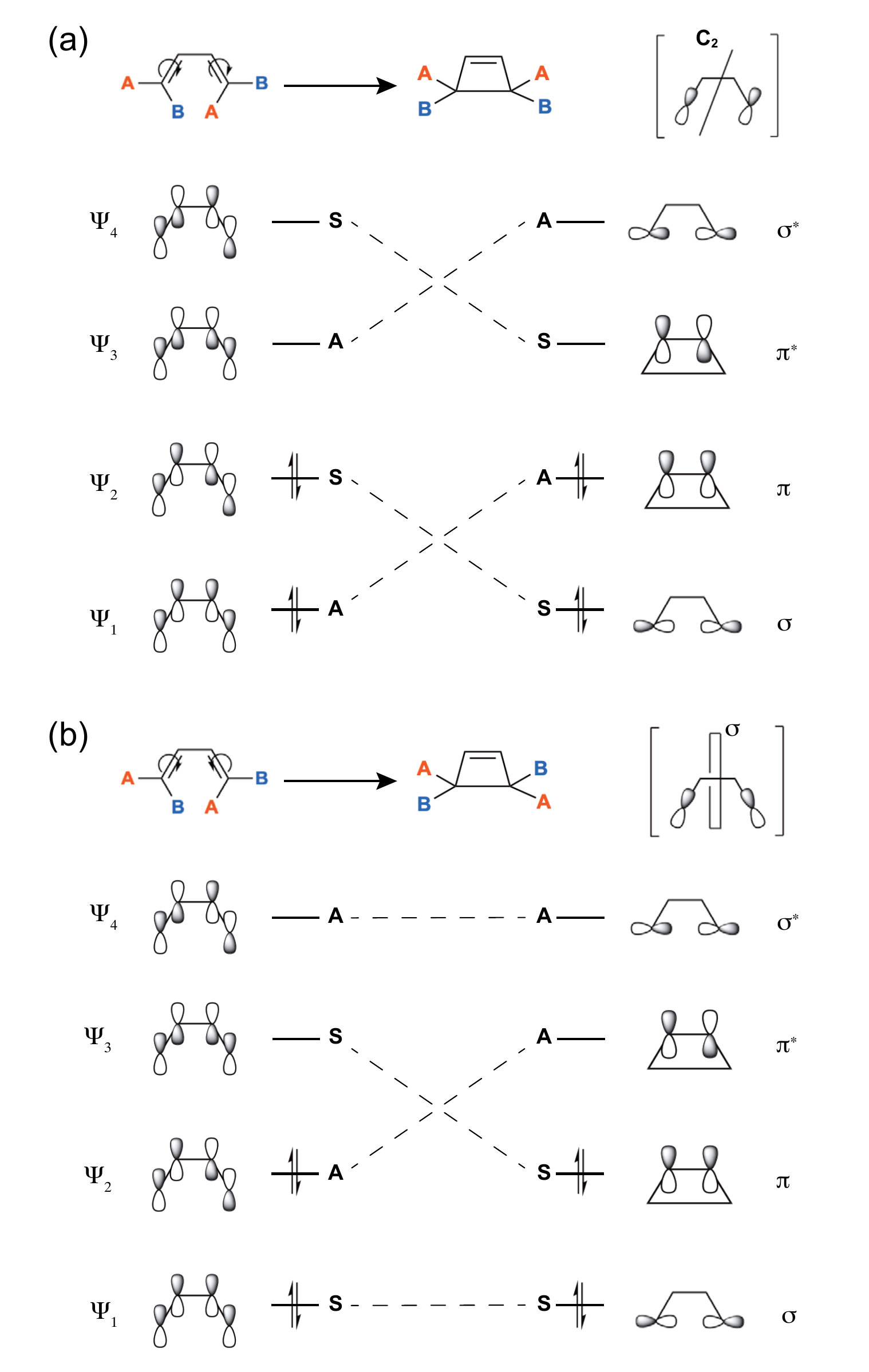}
    \caption{Sketch of the molecular orbital correlation diagram of the (a) conrotatory pathway and (b) disrotatory pathway. The conserved symmetries are highlighted in the top right panel of each schematic.}
    \label{fig:MOsCorrelationDiagram}
\end{figure}
This phenomenology is general~\cite{gurarie2011single}. In the presence of strong multi-reference correlations, poles are replaces by zeros close to $\omega = 0$.
Therefore, is vital to not only study the evolution of Green's function poles, but also of the zeros, in particular when studying HOMO/LUMO crossings along a reaction coordinate.

To exemplify this approach, we turn to the $4\pi$ electrocyclization of butadiene to cyclobutene~\cite{Woodward1969-ds}.
Electrocyclizations are ideal model systems, as they display a complex interplay of symmetry, energy level crossings and electronic correlations,
documented by an extensive body of experimental and computational data. Further, these systems are amendable to high-level quantum chemical calculations, enabling a systematically improvable theoretical approach that could be compared to experiments.
The electrocyclization can proceed along two pathways. The conrotatory pathway [Fig.~\ref{fig:MOsCorrelationDiagram}a] is thermally allowed. A $C_2$ symmetry is preserved along the reaction pathway and there is no crossing between unoccupied and occupied MOs along the reaction coordinate.
On the other hand, the disrotatory pathway is thermally forbidden. A mirror symmetry $\sigma$ is preserved along the pathway, which protects a crossing between the HOMO and LUMO of opposite parity [Fig.~\ref{fig:MOsCorrelationDiagram}b].
In the absence of electronic interactions, this distinction would allow for the formulation of topological descriptors of these crossings, as no symmetry allowed terms can be added to the Hamiltonian to gap it out~\cite{Muechler2020-de}. 
But, just as it is the case for cyclobutadiene, the presence of such a crossings implies that electron-electron interactions cannot be neglected, as these two orbitals are half-filled and highly-susceptible to correlation effects due to the lack of Fermi-liquid like screening in molecules. 
In a non-interacting picture, three degenerate ground states with $S = 0$ exist. This degeneracy is lifted upon inclusion of interactions, and the size of the gap depends on the strength of the interactions~\cite{Muechler2014-nq}. 
\begin{figure}[t]
    \includegraphics[width=0.9\columnwidth]{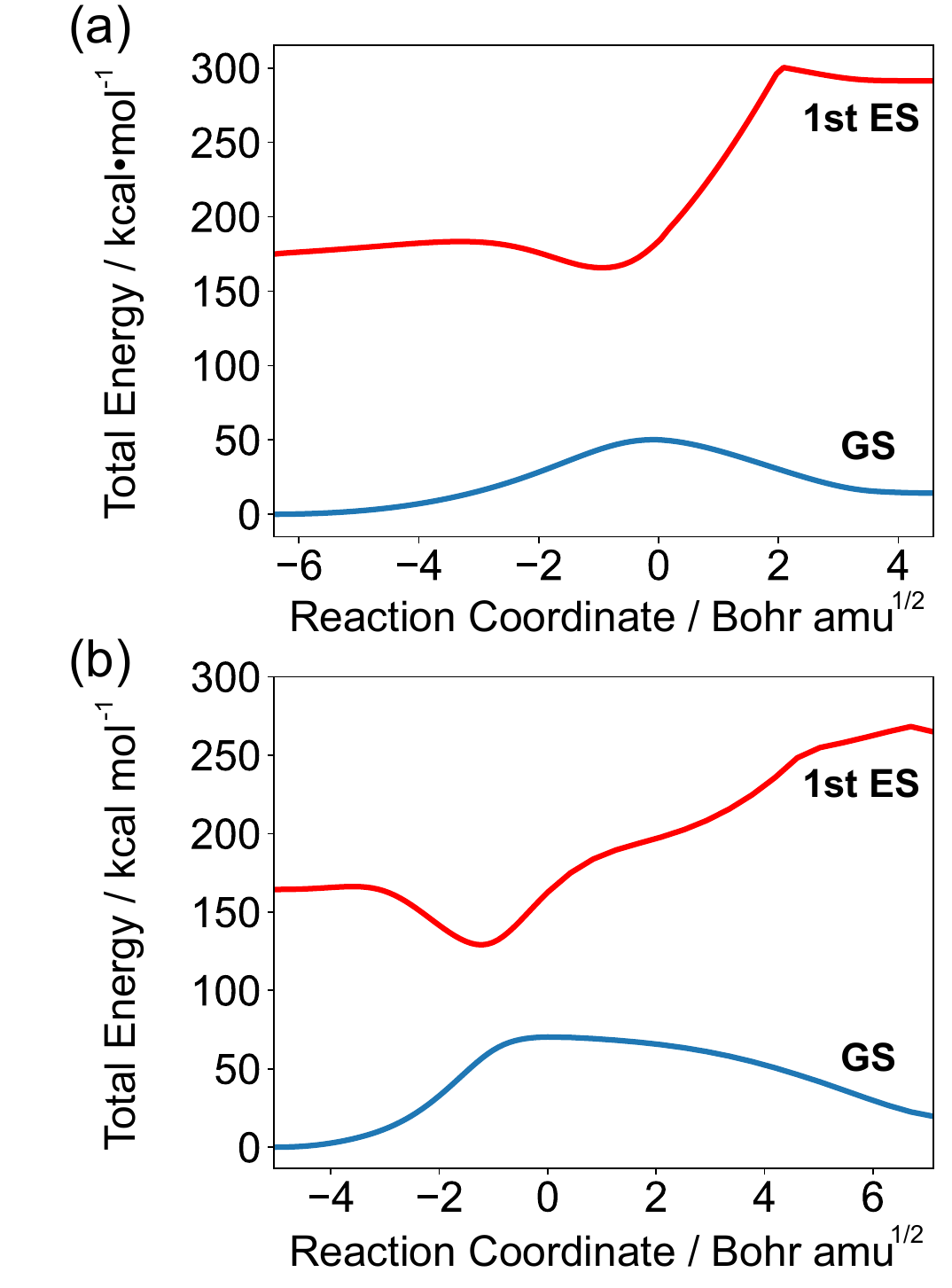}
    \caption{Ground state and first excited state reaction coordinate plot at the CASSCF(4,4)/def2-SVP level. (a) The symmetry-allowed conrotatory ring closure of butadiene. (b) The symmetry forbidden disrotatory ring closure. 
    }
    \label{fig:PESforGSand1stES}
\end{figure}
Figure~\ref{fig:PESforGSand1stES} shows the ground and first excited state for both pathways calculated at the CASSCF(4,4)/def2-SVP level based on intrinsic reaction coordinate (IRC) calculations~\cite{fukui1981path, fukui1970formulation}. 
Both pathways feature a unique ground state, while the barrier of the conrotatory pathway is significantly lower than the disrotatory pathway in accordance with previous calculations~\cite{Longuet-Higgins1965-xc,mirzanejad2025,Zhou2024-ji}. 
\begin{figure*}[htp]
    \includegraphics[width=1.6\columnwidth]{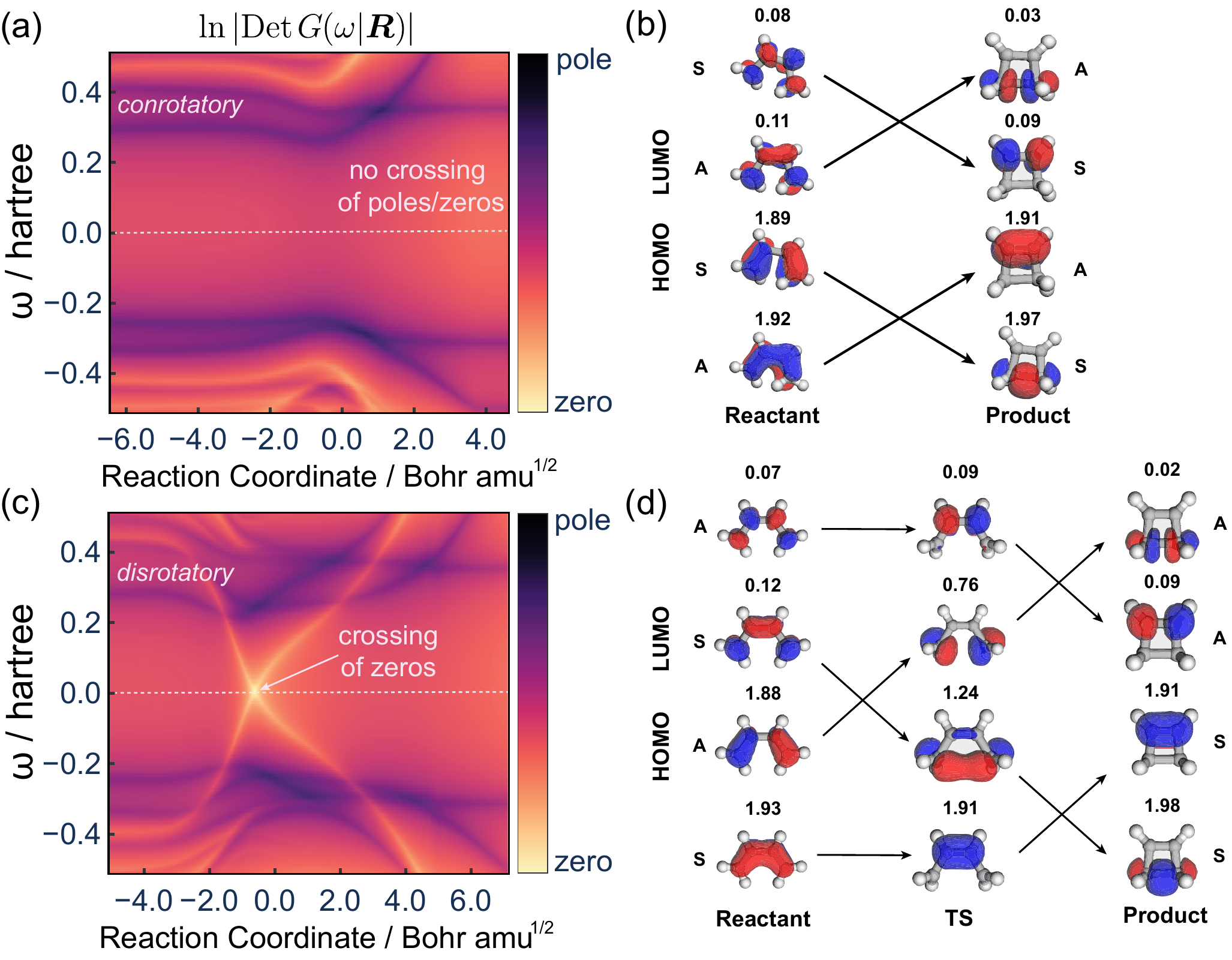}
    \caption{(a) Green's function  plotted as $\ln|\text{Det }G\left(\omega | \bs{R}\right)|$ for the symmetry allowed reaction of butadiene to cyclobutene. (b) Natural orbitals and natural orbital occupation numbers of reactants and products. (c) Green's function plotted as $\ln|\text{Det }G\left(\omega | \bs{R}\right)|$ for the symmetry forbidden reaction of butadiene to cyclobutene. (d) Natural orbitals and natural orbital occupation numbers of the reactant, transition state (TS) and product. The Green's functions are obtained from CASSCF(4,4)/def2-SVP calculations.
    }
    \label{fig:GF for two reaction pathways}
\end{figure*}
This phenomenology cannot be faithfully described by single reference methods such as DFT and even CCSD~\cite{Bulik2015-xf,Monino2022-cu,Zhou2024-ji,mirzanejad2025}. As a result, MO energies are not accessible from conventional ground state calculations, and the distinction between the two pathways, one with a HOMO/LUMO crossing, the other without, is lost.
As we will show below, this information is contained in the Green's function. Analyzing the conrotatory pathway, we find no crossings of poles between occupied and unoccupied states, in accordance with the MO picture.
We find crossings of poles for $\omega < 0$, which correlate with changes in natural orbital occupation numbers (NOONs) within the active space [Fig.~\ref{fig:GF for two reaction pathways}(b)]~\cite{Bright_Wilson1972-mp}. In addition, there are no zeros close to $\omega = 0$, indicating that static correlations do not dominate this reaction.

In contrast to the predictions of MO theory [Fig.~\ref{fig:MOsCorrelationDiagram}b], the Green's function for the symmetry forbidden pathway  does not show a crossing of poles close to the transition state geometry [Fig~\ref{fig:GF for two reaction pathways}c].
Rather, a crossing of zeros at $\omega = 0$ is clearly visible before the transition state, indicating strong static correlation for geometries close to it. Analysis of the natural orbitals shows the exchange of HOMO and LUMO after the crossing of zeros [Fig.~\ref{fig:GF for two reaction pathways}d]. After the TS, we find crossings of poles within the occupied and unoccupied region, which similarly correlate with the change of NOONS. The crossings of zeros occurs at the geometry at which the NOONs of the HOMO and LUMO become degenerate.
These results demonstrate that the crossings of MOs in a single-particle picture are replaced by crossings of zeros when electronic correlations are properly taken into account~\cite{gurarie2011single}.
An analysis based on Green's functions provides both a qualitative and quantitative generalization of orbital crossings between occupied and unoccupied states during chemical reactions, overcoming limitations of MO theory. This approach also allows for the definition of topological invariants associated with these crossings, accounting for both the poles and zeros of the Green's function.

\subsection{Topological invariants of Symmetry-Allowed and Symmetry-Forbidden Electrocyclic Reactions}

In the absence of multi-reference correlations, one can diagnose and characterize crossings using symmetry indicators~\cite{Bradlyn2017-zu,watanabe2017structure,Po2017-gw,Chiu2014-lz}.
In the case of a symmetry forbidden reaction, in which the crossings of single-particle states are protected by a symmetry, one can characterize these states as even or odd under the symmetry. 
The number of occupied single particle states with belonging to the even or odd subspaces must change after the crossing.
One can therefore diagnose crossings by comparing the number of occupied states in each subspace for different molecular geometries, e.g. between the reactants and products. Depending on the details of the system, it is then possible to diagnose the number of crossings, or at least their parity.
In the presence of strong correlations, i.e. when poles and zeros cross $\omega = 0$, this classification is no longer valid~\cite{gurarie2011single}. Instead a topological invariant taking into account both poles and zeros has to be defined.
Such an invariant is given as~\cite{gurarie2011single,Zhou2020-zf}
\begin{equation}
    N (\bs{R}) =\oint_C \frac{d \omega}{2 \pi i} \frac{\partial \ln \operatorname{Det } G\left(\omega | \bs{R}\right)}{\partial \omega}
    \label{equ:topological invariant formula}
\end{equation}
where $C$ denotes an integration path in the complex $\Omega$ plane, which we choose as the complex upper half plane. The topological invariant counts the number of zeros and poles on the real line contained within the plane, i.e. poles and zeros for $\omega > 0$ for our choice of plane. The invariant is calculated for a specific molecular geometry $\bs{R}$. The invariant is ill-defined if a pole or zero is present at $\omega=0$.
\\

It is therefore meaningful to calculate the difference between of invariants $\Delta N = N(\bs{R}_i) - N(\bs{R}_j)$ between different molecular geometries $\bs{R}_i$ and $\bs{R_j}$. If the electronic Hamiltonian preserves the number of electrons the difference of invariants vanishes $\Delta N \equiv 0$, which means that for each pole (zero) that crosses $\omega = 0$ from below is compensated by a pole (zero) that crosses $\omega = 0$ from above as one changes the molecular geometry from $\bs{R}_i$ to $\bs{R}_j$. This result reflects charge conservation, as it ensures that the number of electrons in the system stays constant between the two chosen geometries.\\

However, in presence of spatial symmetries one can define an invariant for each symmetry sector. For the conrotatory pathway, this symmetry is a $C_2$ rotation, while it is a mirror symmetry for the disrotatory pathway. The CAS-orbitals transform as irreducible representations of these  symmetries, and therefore allow us to block-diagonalize the Green's function at each molecular geometry as $G\left(\omega | \bs{R}\right) = G^+\left(\omega | \bs{R}\right) \oplus  G^-\left(\omega | \bs{R}\right)$, thus defining invariants $N_+(\bs{R})$ and $N_-(\bs{R})$ (See SI for details).\\

The condition $\Delta N = 0$ implies that
\begin{equation}
\label{eq:invariant_constraint_sym}
\begin{split}
    \Delta N_+ + \Delta N_- &= 0 \\ 
    \Delta N_+ - \Delta N_- &\neq 0.
\end{split}
\end{equation}
 A change of $\Delta N_+ - \Delta N_-$ therefore measures a switching of poles (zeros) between two different molecular geometries that belong to different blocks of $G\left(\omega | \bs{R}\right)$.
 Due to Eq.~\ref{eq:invariant_constraint_sym}, it is sufficient to compute the change of one block, i.e
 $\Delta N_+$ or $\Delta N_-$.\\

\begin{figure}[ht]
    \includegraphics[width=0.99\columnwidth]{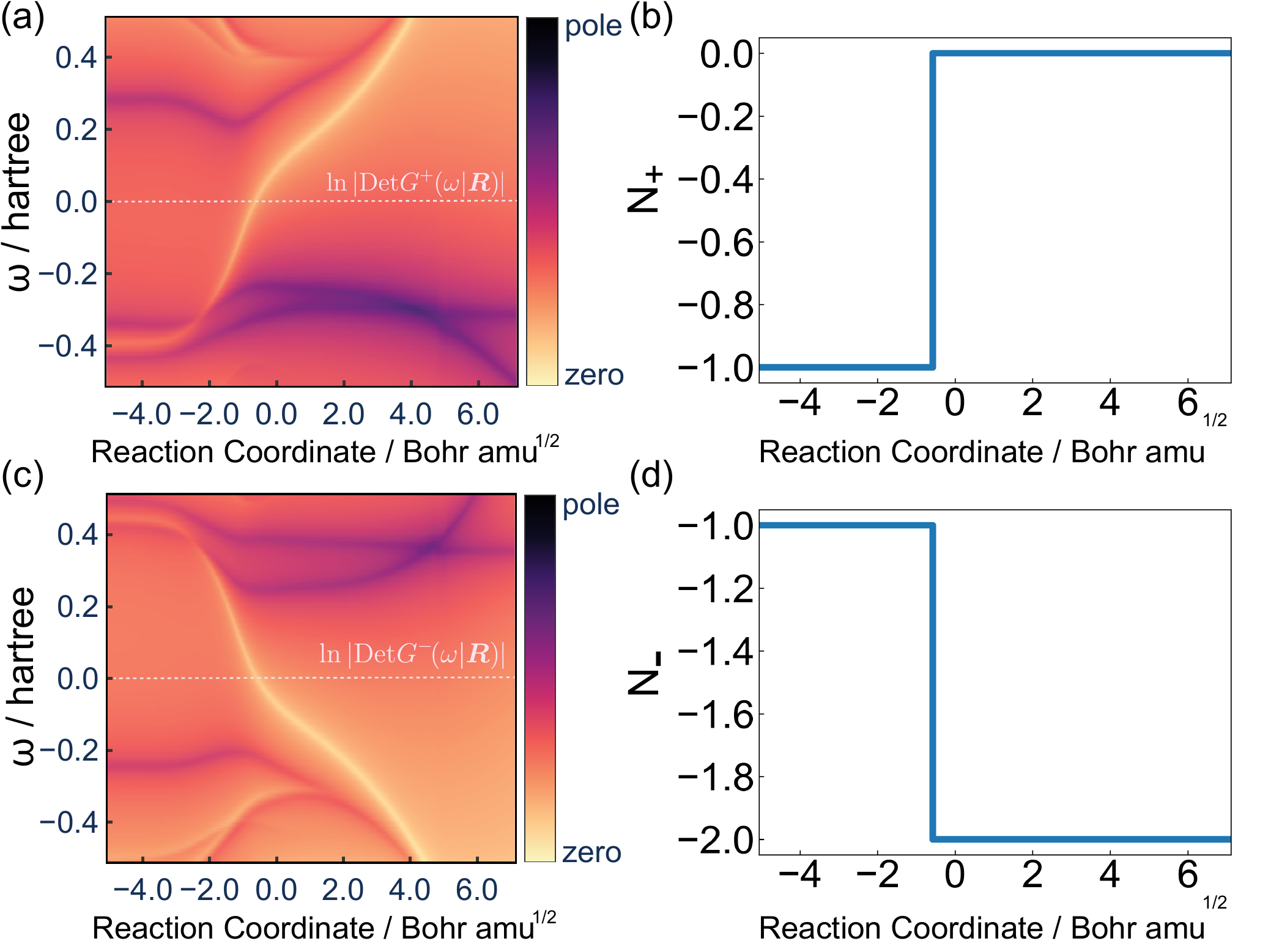}
   \caption{$\ln|\text{Det }G^{\pm}\left(\omega | \bs{R}\right)|$ and topological invariants along the reaction coordinate for the symmetry forbidden reaction for the $\sigma_v$ even (a \& b) and odd (c \& d) subspaces.}
    \label{fig:GF plus and minus part (fobidden)}
\end{figure}

\begin{figure}[ht]
    \includegraphics[width=0.99\columnwidth]{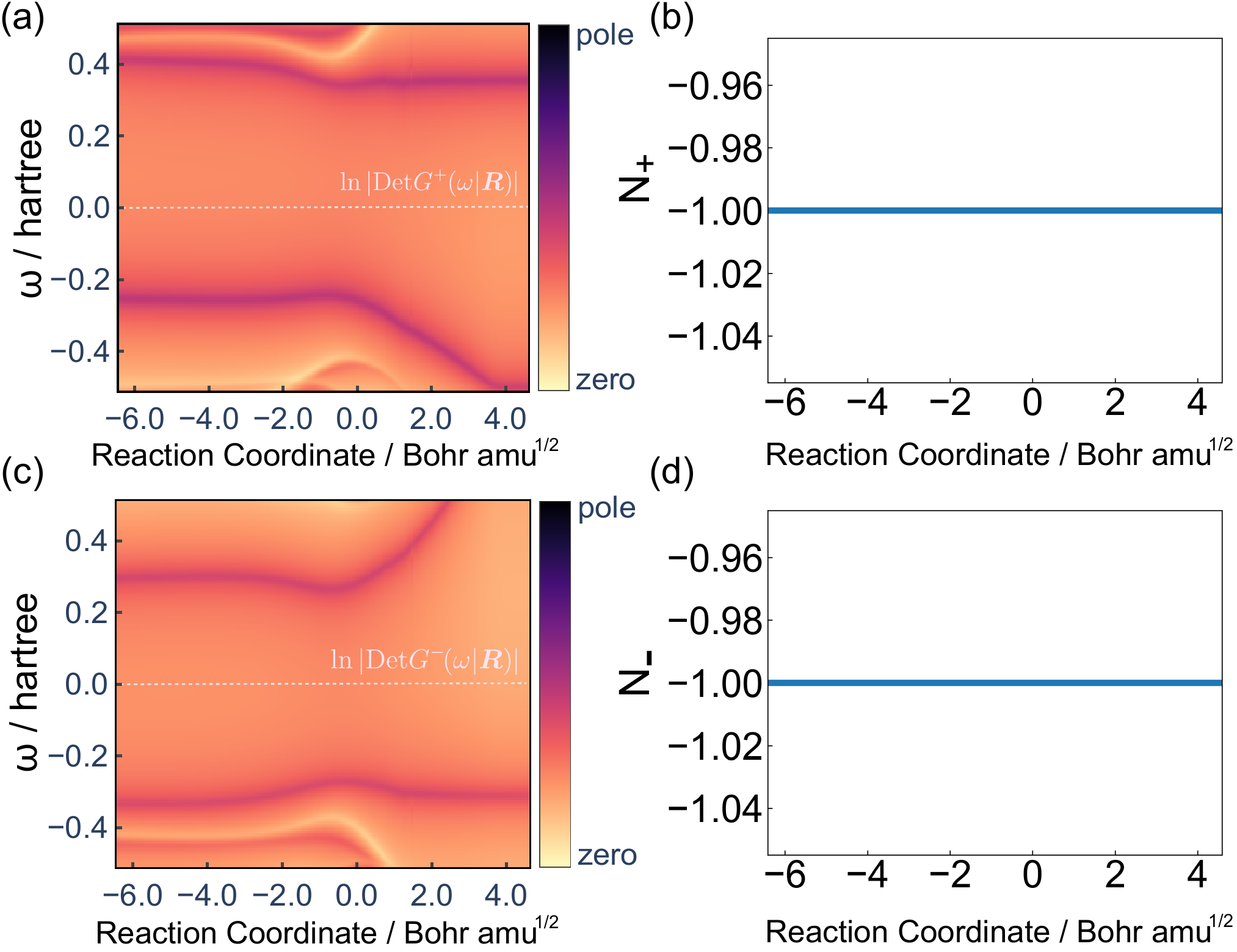}
    \caption{$\ln|\text{Det }G^{\pm}\left(\omega | \bs{R}\right)|$ and topological invariants along the reaction coordinate for the symmetry allowed reaction for the $C_2$ even (a \& b) and odd (c \& d) subspaces.
   }
    \label{fig:GF plus and minus part (allowed)}
\end{figure}

Figure~\ref{fig:GF plus and minus part (fobidden)} plots $G^+(\omega|\bs{R})$ and $G^{-}(\omega|\bs{R})$ for the symmetry forbidden pathway as well as $N_+$/$N_-$ along the reaction coordinate. According to Eq. \ref{equ:topological invariant formula}, $N_+$ and $N_-$ can only change as poles or zeros cross $\omega = 0$, while they are ill-defined at the crossing point. A zero moves from negative to positive ($G^+(\omega|\bs{R})$) and from positive to negative ($G^-(\omega|\bs{R})$), which results in $\Delta N_+ = 1$ comparing geometries after and before the crossing point.
For the symmetry allowed reaction (Fig.~\ref{fig:GF plus and minus part (allowed)}), we find no poles and zeros crossing $\omega = 0$ and therefore no change of $\Delta N_+$ along the reaction path between different molecular geometries.

\section{Discussion}
This distinction via invariants demonstrates the utility of a topological approach for reactions in which symmetry is conserved along the reaction pathway. To determine whether a reaction is allowed or forbidden, it is not necessary to compute the full reaction pathway. Instead, it is sufficient to evaluate $\Delta N_+$ or $\Delta N_-$ between reactants and products. 

\begin{figure}[t]
    \includegraphics[width=0.85\columnwidth]{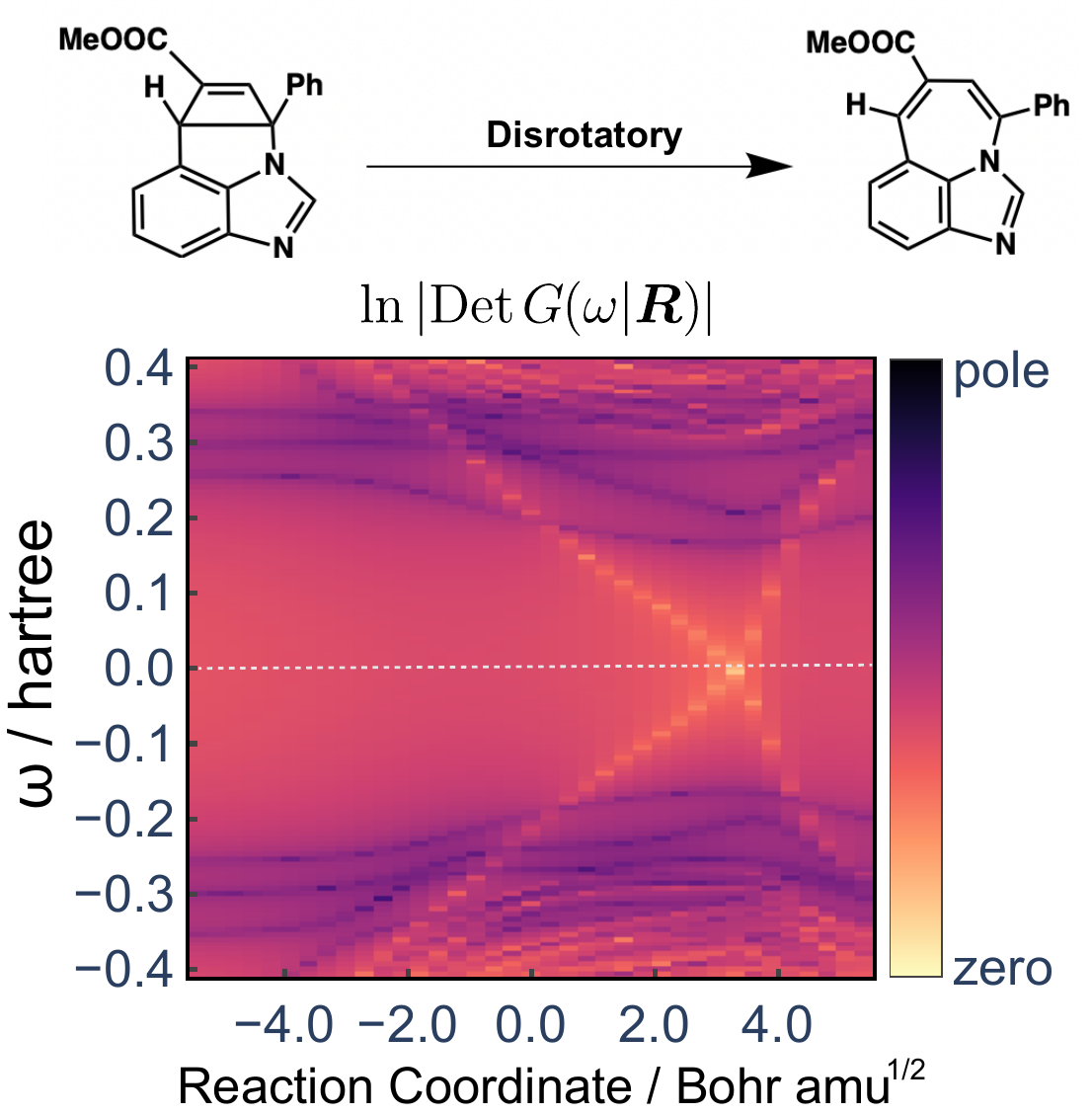}
    \caption{$\ln|\text{Det }G\left(\omega | \bs{R}\right)|$ of a reaction proposed in ref.~\citep{mirzanejad2025} along the forbidden, disrotatory pathway. The molecule does not possess any symmetries along this pathway, yet the crossing of Green's function zeros remains well defined. The Green's function is obtained at the CASSCF(8,8)/def2-SVP level.}
    \label{fig:GF for symmetry breaking reaction}
\end{figure}
The crossings of poles and zeros, as well as the definition of the invariants $N_+$ and $N_-$ is contingent on the presence of a spatial symmetry along the reaction path. While this is the case for the canonical WHR reactions discussed above, most molecules of relevance to chemical problems break these symmetries through substituents, which, to varying degree, affect the orbitals involved in the reaction.
Symmetry breaking introduces an energy scale $\Delta$, which must be compared to the single-particle energy scale $E_{sp}$ given by the off-diagonal matrix elements of the single-particle Hamiltonian. When $\Delta \ll E_{sp}$, the single-particle description remains valid and the results of the previous section hold even without electron–electron interactions. By contrast, for $\Delta \sim E_{sp}$ or larger, one must also consider the interaction energy $U$. In this regime, the ratio $\frac{\Delta}{U}$ controls how symmetry breaking competes with correlation effects.  
To quantify these scales for a realistic reaction, we applied our formalism to a recently proposed a planar 4$\pi$ photoswitch [Fig.~\ref{fig:GF for symmetry breaking reaction}].
For this molecule, the ring-opening reaction is predicted to proceed along a single-step disrotatory reaction pathway, while the all spatial symmetries are broken~\cite{mirzanejad2025}. 
We find that the crossing of zeros at $\omega=0$ is maintained, despite the explicit breaking of symmetries. These results suggest that the symmetry breaking energy scale induced by the asymmetric substituents is lower than the energy scale of static correlation, and validates the applicability of our approach.

The ideas presented here can be generalized to reactions that involve different spin states, e.g. by defining Green's functions and invariants for different spin sectors of the Hamiltonian. This approach would allow to define rigorously spin-dependent Berry-forces beyond mean-field approximations~\cite{Bian2022-vt}, as generalizations of the Berry curvature can be calculated from the Green's function for each spin-sector~\cite{Kapustin2020-xh,Chen2022-ab}. Further, a connection between the Green's function zeros and features of the PES such as second order saddle points might be attainable~\cite{mirzanejad2025}.

A central result of topological band theory is that Hamiltonians with different topological invariants are not adiabatically connected. Having established a connection between topological band theory and orbital symmetry controlled reactions, this suggests that our approach might be able to diagnose nonadiabatic effects that have previously been highlighted in orbital-symmetry controlled reactions by Butler\textit{ et. al.}~\cite{Waschewsky1994-er,Butler1998-ek} as well as \citet{Morihashi1982-ib}.


\section{Methods}
All calculations have been performed using PySCF and using the CASSCF method~\cite{sun2007python, sun2020recent}.
The Green's functions are calculated via the TRIQS package, using one- and two-electron integrals from the converged CASSCF active space orbitals~\cite{parcollet2015triqs}. Details about the active space orbitals, and technical details of the Green's function implementation can be found in the SI.

\begin{acknowledgments}
We thank Raquel Queiroz for discussions, in particular from bringing Ref.~\citenum{gurarie2011single} to our attention.
This work was supported by the U.S. Department of Energy, Office of Science, Office of Basic Energy Sciences, CPIMS program, under Award DE-SC0025352.
\end{acknowledgments}
\bibliography{lit}

\clearpage


\setcounter{equation}{0}
\setcounter{figure}{0}
\setcounter{table}{0}
\setcounter{section}{0}
\setcounter{page}{1}
\makeatletter
\renewcommand{\theequation}{S\arabic{equation}}
\renewcommand{\thefigure}{S\arabic{figure}}
\onecolumngrid
\section{Details of First Principle Calculations }
\subsection{Computational methods}

We utilize the GAMESS program for geometry optimizations and frequency calculations at the CASSCF/def2-SVP level~\cite{GAMESS}. Transition structures are confirmed via intrinsic reaction coordinate (IRC) calculations. To calculate the total energy of the first excited state we used state-averaged CAS for two states.

To compute the Green's functions, we use PySCF to calculate the active space one- and two-electron integrals in the CAS active space ($h_{P Q}$ and $g_{P Q R S}$ in Eq.~~\ref{equ:one-electron integral})~\cite{Sun2020-lo}.
We used the natural orbitals from unrestricted second order Møller–Plesset perturbation theory (MP2) as a initial guess for the CAS-orbitals at the stationary points for each reaction. Subsequent calculations along the IRC towards reactants or products used interpolated CAS-orbitals from the previous geometry as initial guess. Custom Python scripts were used to generate plots and visualize orbitals.\\

After completing the CASSCF calculation, we construct the Hamiltonian of the active space. The definition of many-electron Hamiltonian in second quantized form~\cite{Helgaker2000-vr}: 

\begin{equation}
\hat{H}=\sum_{P Q} h_{P Q} a_P^{\dagger} a_Q+\frac{1}{2} \sum_{P Q R S} g_{P Q R S} a_P^{\dagger} a_R^{\dagger} a_S a_Q - \mu\sum_{P}a_P^\dagger a_P
\end{equation}

\begin{equation}
h_{P Q}=\int \phi_P^*(\mathbf{x})\left(-\frac{1}{2} \nabla^2-\sum_I \frac{Z_l}{r_I}\right) \phi_Q(\mathbf{x}) \mathrm{d} \mathbf{x}
\label{equ:one-electron integral}
\end{equation}

\begin{equation}
g_{P Q R S}=\iint \frac{\phi_P^*\left(\mathbf{x}_1\right) \phi_R^*\left(\mathbf{x}_2\right) \phi_Q\left(\mathbf{x}_1\right) \phi_S\left(\mathbf{x}_2\right)}{r_{12}} \mathrm{~d} \mathbf{x}_1 \mathrm{~d} \mathbf{x}_2
\label{equ:two-electron integral}
\end{equation}

\begin{equation}
\mu = -\frac{IP+EA}{2}
\label{equ: chemical potential}
\end{equation}

$Z_I$ is the nuclear charge of atom $I$, $r_I$ is the relative  distance between electron-nucleus, $r_{12}$ is the electron-electron distance and $R_{IJ}$ is the internuclear separation. $P, Q, R, S$ represent the index of orbitals included in the active space. And we adopt the definition of the Mulliken chemical potential (Eq.~~\ref{equ: chemical potential}), where IP and EA represent ionization potential and electron affinity, respectively. The molecular one-electron integral (Eq.~~\ref{equ:one-electron integral}), two-electron integral (Eq.~\ref{equ:two-electron integral}), ionization potential and electron affinity are extracted from the previous CASSCF calculation.

\begin{figure}[htb]
    \centering
    \includegraphics[width=0.5\textwidth]{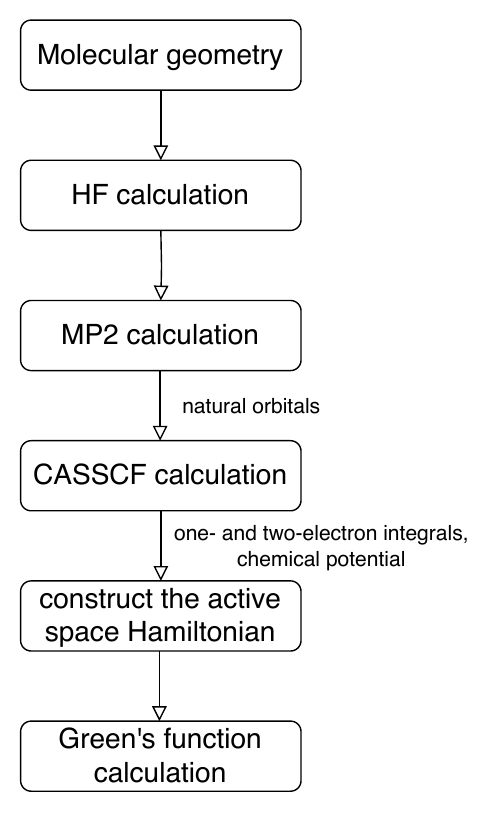}
    \caption{flowchart: Green's function calculations }
    \label{fig:calculation flowchart}
\end{figure}

Subsequently, we employed TRIQS to compute the Green's function along the reaction coordinate~\cite{Parcollet2015-or}. The Green's function heatmap plot displays the frequency($\omega$) versus reaction coordinate, with each point representing $\ln|\text{Det }G\left(\omega | \bs{R}\right)|$ at a specific frequency and reaction coordinate to highlight poles and zeros at the same time.

To obtain the topological invariants for each subspace, we block diagonalize the Green's function into two blocks,
\begin{equation}
G\left(\omega | \bs{R}\right) = G^+\left(\omega | \bs{R}\right) \oplus  G^-\left(\omega | \bs{R}\right)    
\end{equation}
The representations of the unitary matrices for this transformation where obtained from the irreducible representations of the CAS-orbitals through PySCF. For the conrotatory pathway, the relevant symmetry is a $C_2$ symmetry, while for the disrotatory the relevant symmetry is a mirror symmetry $\sigma_v$ as highlighted in Fig.~4 in the main text. 
The corresponding topological invariants were defined as $N_+(\bs{R})$ and $N_-(\bs{R})$, respectively.\\
\begin{figure}[t]
\label{fig-app:integral path}
\includegraphics[width=0.8\columnwidth]{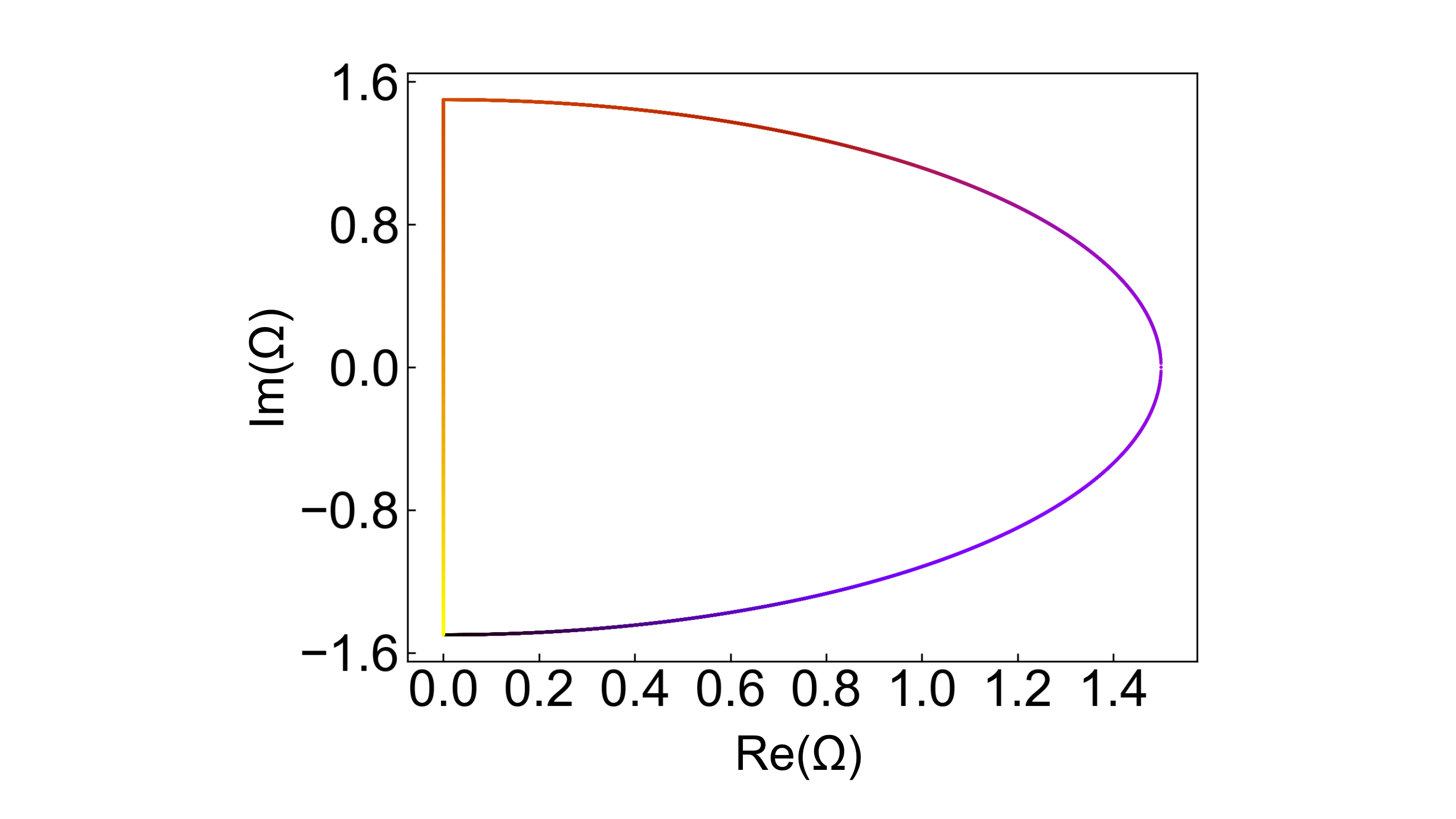}
\caption{Integration contour in the complex plane. The initial point of the contour is indicated in black and the last point is indicated in yellow.}
\end{figure}
To calculate the topological invariant ($N_{+/-}(\bs{R})$), we select a cutoff by integrating $\omega$ within the interval $[-C, C]$. This interval is enclosed by connecting $C$ and $-C$ through a semicircular path in the complex plane, as illustrated in Fig.~\ref{fig:calculation flowchart}. We calculate $\text{Det }G\left(\omega | \bs{R}\right)$ along this curve and plot $\mathrm{Im} \left[  \text{Det } G^{+/-}(\omega|\bs{R})\right ] $ versus $\mathrm{Re} \left[ \text{Det } G^{+/-}(\omega|\bs{R}) \right ] $. The topological invariant (aka winding number) is determined by counting how many times the plot encircles the origin in a clockwise or counterclockwise direction. For particularly complex loops, it can be difficult to evaluate the winding number directly using this method. Therefore, we  use the so-called \textit{crossing rule} to evaluate the winding number, which works as follows:
Initially, the winding number is zero.
Pick a starting point outside the loop and move towards the origin in a straight line. 
If the loop is crossed, and the direction of the curve that is crossed goes from the left to the right, the winding number increases by one.
If direction of the curve crossed is from right to the left, decreases the winding number by one.
For more details about crossing rules, we refer to Ref.~\citenum{Needham2023-kr}.

\begin{figure}[ht]
\label{fig-app:winding_number_butadiene_to_cyclobutene_allowed}
\includegraphics[width=0.99\textwidth]{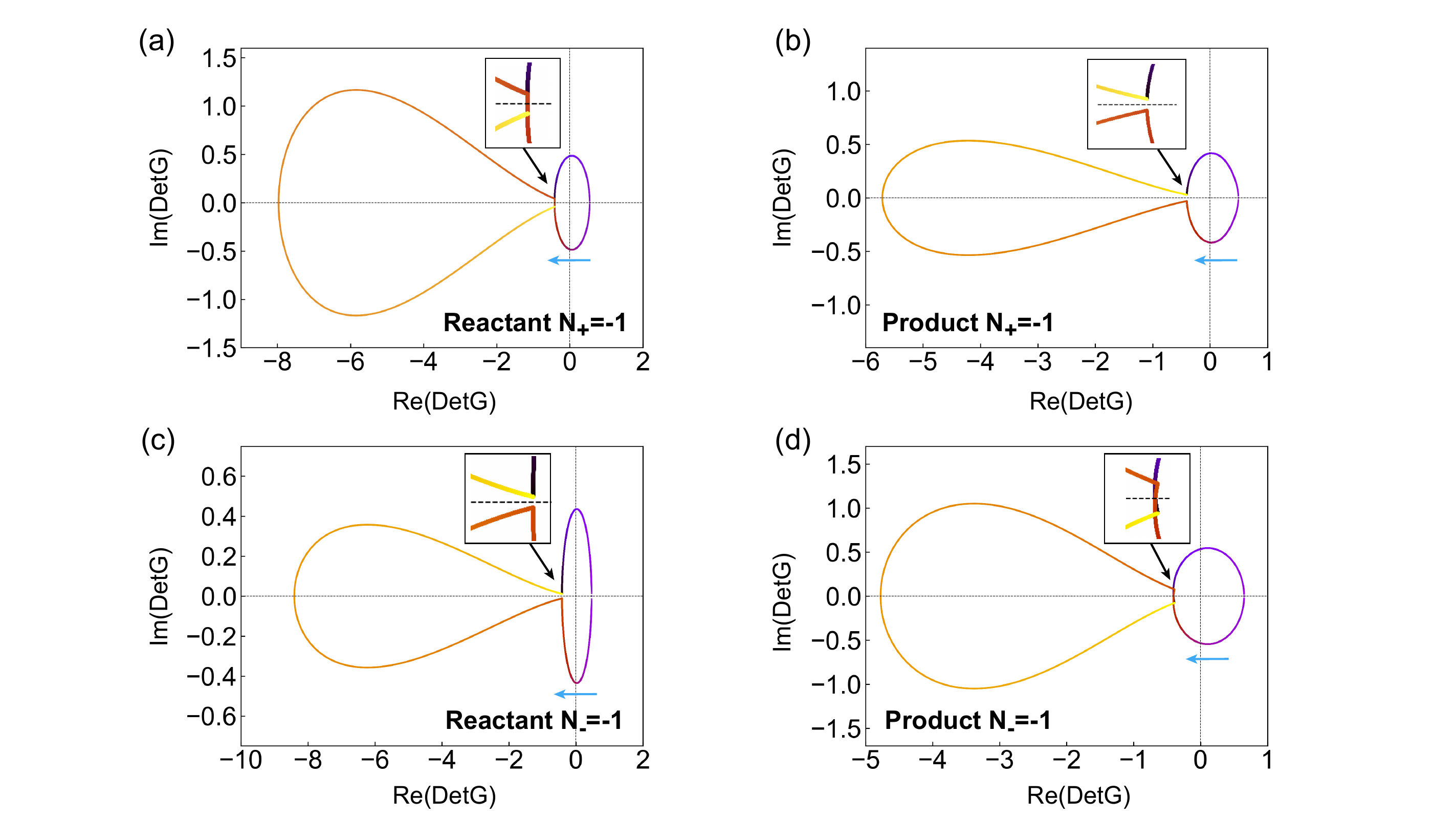}
\caption{Winding numbers of the conrotatory, symmetry allowed reaction for reactant and product geometries. $G^{+}(\omega|\bs{R})$ part: (a) reactant, (b) product. $G^{-}(\omega|\bs{R})$ part: (c) reactant, (d) product.}
\end{figure}

\begin{figure}[ht]
\includegraphics[width=0.99\textwidth]{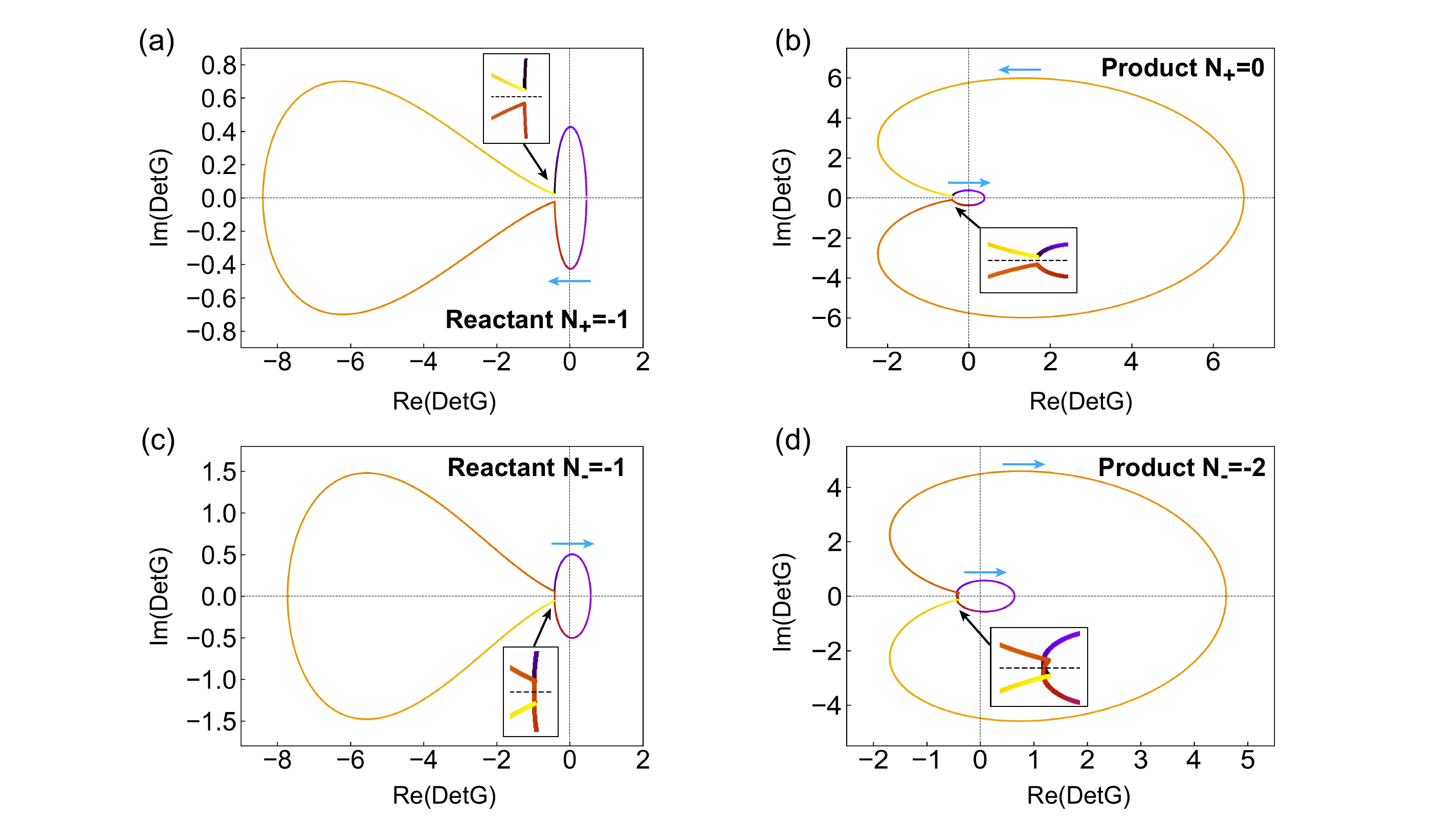}
\caption{Winding numbers of the disrotatory, symmetry forbidden reaction for reactant and product geometries. $G^{+}(\omega|\bs{R})$ part: (a) reactant, (b) product. $G^{-}(\omega|\bs{R})$ part: (c) reactant, (d) product.}
\label{fig-app:winding_number_butadiene_to_cyclobutene_forbidden}
\end{figure}
\clearpage

\subsection{Cartesian Coordinates and active space orbitals}
\subsubsection{Butadiene to Cyclobutene}

\begin{table}[ht]
\centering
\caption{Cartesian coordinates of optimized TS for Butadiene to Cyclobutene (symmetry allowed pathway) calculated at the CASSCF(4,4)/def2-SVP level of theory}
\label{tab:tab2}
\begin{tabular}{|c|c|c|c|}
\hline
Atomic Symbol & x & y & z\\\hline
C & -3.09698439 & 1.17951632 & 0.16758497\\
C & -2.06908488 & 0.52281088 & -0.44725546\\
C & -3.20352459 & 0.57596689 & 1.49097860\\
C & -1.30338478 & -0.09057832 & 0.60867476\\
H & -1.87494135 & 0.41670525 & -1.50920999\\
H & -0.83910215 & -1.06939566 & 0.49497187\\
H & -0.79861146 & 0.55856788 & 1.30909038\\
H & -3.69765520 & 2.00121927 & -0.20634829\\
H & -3.44916940 & 1.16169369 & 2.37621284\\
H & -3.49097323 & -0.46342605 & 1.55438280\\
\hline
\end{tabular}
\end{table}

\begin{table}[H]
\centering
\caption{Cartesian coordinates of product for Butadiene to Cyclobutene (symmetry allowed pathway) calculated at the CASSCF(4,4)/def2-SVP level of theory}
\label{tab:tab2}
\begin{tabular}{|c|c|c|c|}
\hline
Atomic Symbol & x & y & z\\\hline
C  &   -3.17146349  &   1.11432397  &   0.06286150\\
C  &   -2.03004956  &   0.65461206  &  -0.47479331\\
C  &   -2.91020942  &   0.47780326  &   1.44256830\\
C  &   -1.56586814  &  -0.05285020  &   0.77611250\\
H  &   -1.58325458  &   0.74785513  &  -1.45692015\\
H  &   -1.49369264  &  -1.14043438  &   0.71065664\\
H  &   -0.65249038  &   0.33375841  &   1.23284853\\
H  &   -3.98674035  &   1.71433961  &  -0.32127821\\
H  &   -2.76802850  &   1.17809510  &   2.26703858\\
H  &  -3.61133671   & -0.30175376   &  1.74432707\\
\hline
\end{tabular}
\end{table}

\begin{table}[h]
\centering
\caption{Cartesian coordinates of reactant for Butadiene to Cyclobutene (symmetry allowed pathway) calculated at the CASSCF(4,4)/def2-SVP level of theory}
\label{tab:tab2}
\begin{tabular}{|c|c|c|c|}
\hline
Atomic Symbol & x & y & z\\\hline
C  &   -3.09340835  &   1.13042045  &   0.34887734\\
C  &   -1.98993051  &   0.41447654  &  -0.32534304\\
C  &   -3.62827516  &   0.80390429  &   1.54641604\\
C  &   -0.95083332  &  -0.18146992  &   0.28743789\\
H  &   -2.03468037  &   0.39101189  &  -1.40977716\\
H  &   -0.18817978  &  -0.69780713  &  -0.28123748\\
H  &   -0.82970405  &  -0.15841198  &   1.36363339\\
H  &   -3.50671363  &   1.97792351  &  -0.18947060\\
H  &   -4.42966700  &   1.39007092  &   1.97846949\\
H  &   -3.28690767  &  -0.05470822  &   2.11210203\\
\hline
\end{tabular}
\end{table}
\begin{table}[h]
\centering
\caption{Cartesian coordinates of optimized TS for Butadiene to Cyclobutene (symmetry forbidden pathway) calculated at the CASSCF(4,4)/def2-SVP level of theory}
\label{tab:tab2}
\begin{tabular}{|c|c|c|c|}
\hline
Atomic Symbol & x & y & z\\\hline
C  &   -0.66840178  &  -0.66103232  &  -0.00818259\\
C  &    0.67361021  &  -0.66141027  &  -0.00862591\\
C  &   -1.43289435  &   0.60805488  &   0.03538374\\
C  &    1.44137657  &   0.60566473  &   0.03363175\\
H  &    1.21823382  &  -1.61539078  &  -0.04042951\\
H  &    2.14264894  &   0.78695387  &   0.83953464\\
H  &    1.58068931  &   1.18977380  &  -0.86852658\\
H  &   -1.21529257  &  -1.60057521  &  -0.03935051\\
H  &   -2.13548136  &   0.78888953  &   0.83994591\\
H  &   -1.56158030  &   1.19736910  &  -0.86460614\\
\hline
\end{tabular}
\end{table}
\begin{table}[h]
\centering
\caption{Cartesian coordinates of product for Butadiene to Cyclobutene (symmetry forbidden pathway) calculated at the CASSCF(4,4)/def2-SVP level of theory}
\label{tab:tab2}
\begin{tabular}{|c|c|c|c|}
\hline
Atomic Symbol & x & y & z\\\hline
C  &   -0.66984105  &  -0.79434568  &  -0.02512581\\
C  &    0.67388409  &  -0.79554427  &  -0.02875960\\
C  &   -0.79400730  &   0.71229649  &   0.04336154\\
C  &    0.80414504  &   0.71111393  &   0.04107838\\
H  &    1.41726863  &  -1.59610069  &  -0.06718586\\
H  &    1.26075172  &   1.11085021  &   0.94924062\\
H  &    1.25962353  &   1.19466817  &  -0.82582450\\
H  &   -1.41343272  &  -1.58071411  &  -0.05936407\\
H  &   -1.24933398  &   1.11309564  &   0.95151293\\
H  &   -1.25149596  &   1.19291830  &  -0.82399887\\
\hline
\end{tabular}
\end{table}
\begin{table}[h]
\centering
\caption{Cartesian coordinates of reactant for Butadiene to Cyclobutene (symmetry forbidden pathway) calculated at the CASSCF(4,4)/def2-SVP level of theory}
\label{tab:tab2}
\begin{tabular}{|c|c|c|c|}
\hline
Atomic Symbol & x & y & z\\\hline
C  &   -0.73492372  &  -0.56231099  &   0.04531593\\
C  &    0.74057925  &  -0.56246090  &   0.04784043\\
C  &  -1.54649913   &  0.51096493   & -0.02430039\\
C  &    1.55419683  &   0.50651175  &  -0.02794648\\
H  &    1.20860767  &  -1.55056751  &   0.12000136\\
H  &    2.62945938  &   0.38468352  &  -0.01693699\\
H  &    1.18276739  &   1.52091098  &  -0.10236876\\
H  &   -1.19537926  &  -1.54244065  &   0.10699496\\
H  &   -2.62218237  &   0.39323509  &  -0.01903062\\
H  &   -1.17003775  &   1.52419770  &  -0.08757465\\
\hline
\end{tabular}
\end{table}

\subsubsection{Planar 4$\pi$ photoswitch}

\begin{table}[ht]
\centering
\caption{Cartesian coordinates of optimized TS for planar 4$\pi$ photoswitch reaction calculated at the CASSCF(8,8)/def2-SVP level of theory}
\label{tab:tab2}
\begin{tabular}{|c|c|c|c|}
\hline
Atomic Symbol & x & y & z\\\hline
C  &  -0.68646049 &   0.74695677  & -0.26518306\\
C  &  -0.73827797 &  -0.72022313  & -0.19584633\\
C  &  -0.71085125 &  -3.55579734  &  0.31736165\\
C  &  -1.31871355 &  -1.62466264  & -1.08968341\\
C  &  -0.16956536 &  -1.31195188  &  0.93895537\\
C  &  -0.15639506 &  -2.65334058  &  1.24435139\\
C  &  -1.27998984 &  -3.01678538  & -0.83651161\\
H  &  -1.78895831 &  -1.26381147  & -1.99523020\\
H  &  -1.72385740 &  -3.68402672  & -1.56389070\\
H  &  -0.72922039 &  -4.61887884  &  0.51064384\\
H  &  -1.47488511 &   1.25954831  & -0.80355763\\
C  &   0.28464389 &   0.76984793  &  2.05043721\\
C  &   0.65886915 &   1.40411985  & -0.18743424\\
C  &   1.19861233 &   1.39390028  &  1.04021025\\
C  &   1.39995897 &   1.89838302  & -1.38580012\\
O  &   2.46708703 &   2.40970922  & -1.34519541\\
O  &   0.72253424 &   1.70862389  & -2.50057220\\
C  &   1.31340802 &   2.12376046  & -3.71306062\\
H  &   0.60278982 &   1.88054800  & -4.49621964\\
H  &   1.50516236 &   3.19430184  & -3.69939637\\
H  &   2.25073671 &   1.59839034  & -3.88155103\\
H  &   2.21268296 &   1.70370245  &  1.26901102\\
N  &   0.28589794 &  -0.64505559  &  2.04045677\\
N  &   0.33208391 &  -2.81264448  &  2.52984500\\
C  &   0.56166804 &  -1.62683094  &  2.96074963\\
H  &   0.94901824 &  -1.39009523  &  3.94072556\\
C  &  -0.00107836 &   1.45962024  &  3.34928083\\
H  &  -0.21320437 &   2.51322961  &  3.16947246\\
H  &  -0.85843676 &   1.01205695  &  3.85135221\\
H  &   0.85915458 &   1.41035175  &  4.02701950\\
\hline
\end{tabular}
\end{table}

\begin{table}[!ht]
\centering
\caption{Cartesian coordinates of product for planar 4$\pi$ photoswitch reaction calculated at the CASSCF(8,8)/def2-SVP level of theory}
\label{tab:tab2}
\begin{tabular}{|c|c|c|c|}
\hline
Atomic Symbol & x & y & z\\\hline
C   &   0.13194802  &   0.02506223  &  -1.09168410\\
C   &  -0.36576492  &  -1.17663205  &  -0.41850802\\
C   &  -1.32690811  &  -3.60743833  &   0.76794356\\
C   &  -0.98295718  &  -2.17476368  &  -1.16612065\\
C   &  -0.24827439  &  -1.43162608  &   0.97211659\\
C   &  -0.70265645  &  -2.60211253  &   1.54609895\\
C   &  -1.46222675  &  -3.37571764  &  -0.58461285\\
H   &  -1.09523642  &  -2.02926564  &  -2.23253226\\
H   &  -1.93308270  &  -4.11443520  &  -1.21931565\\
H   &  -1.67700648  &  -4.51568985  &   1.23679578\\
H   &  -0.00438553  &   0.03394382  &  -2.16143703\\
C   &   0.86167991  &   0.60631412  &   1.99349451\\
C   &   0.74190885  &   1.10258996  &  -0.53969592\\
C   &   1.04051375  &   1.35354006  &   0.88409495\\
C   &   1.18740273  &   2.21943665  &  -1.44181168\\
O   &   1.70200074  &   3.21299028  &  -1.04911458\\
O   &   0.95879722  &   1.99752009  &  -2.72144365\\
C   &   1.35240567  &   2.99009442  &  -3.64379215\\
H   &   1.08100641  &   2.61502934  &  -4.62530088\\
H   &   0.83624667  &   3.92648840  &  -3.44466448\\
H   &   2.42547798  &   3.15932298  &  -3.59121442\\
H   &   1.48822224  &   2.31871080  &   1.05400848\\
N   &   0.29010919  &  -0.69277030  &   2.01553392\\
N   &  -0.45261529  &  -2.59635425  &   2.89602900\\
C   &   0.12038465  &  -1.47597969  &   3.12541842\\
H   &   0.44914326  &  -1.15470731  &   4.09897423\\
C   &   1.27664340  &   1.14159203  &   3.33980918\\
H   &   1.68955755  &   2.13991475  &   3.22737455\\
H   &   0.42801705  &   1.20167029  &   4.02360106\\
H   &   2.03912401  &   0.51237071  &   3.80258489\\
\hline
\end{tabular}
\end{table}

\begin{table}[!ht]
\centering
\caption{Cartesian coordinates of reactant for planar 4$\pi$ photoswitch reaction calculated at the CASSCF(8,8)/def2-SVP level of theory}
\label{tab:tab2}
\begin{tabular}{|c|c|c|c|}
\hline
Atomic Symbol & x & y & z\\\hline
C  &   -0.46316966 &   0.80869728  &   0.27256081\\
C  &   -0.70543504 &  -0.65206575  &  -0.07928471\\
C  &   -0.67484796 &  -3.52020717  &   0.21292190\\
C  &   -1.18891037 &  -1.44962764  &  -1.10089016\\
C  &   -0.24345320 &  -1.36385739  &   1.01594305\\
C  &   -0.17524883 &  -2.69788671  &   1.26026309\\
C  &   -1.16230154 &  -2.87462640  &  -0.92405868\\
H  &   -1.57645857 &  -1.03834307  &  -2.02346087\\
H  &  -1.54599094  & -3.48445034   & -1.73209500\\
H  &   -0.68561327 &  -4.59903717  &   0.28033912\\
H  &  -1.31754351  &  1.47265327   &  0.13959162\\
C  &    0.19265704 &   0.75756383  &   1.80498528\\
C  &    0.89137906 &   1.38201642  &  -0.09673298\\
C  &    1.43442523 &   1.31957614  &   1.13257849\\
C  &    1.46329761 &   1.83212495  &  -1.38748384\\
O  &    2.55850339 &   2.26127553  &  -1.51424849\\
O  &    0.60081214 &   1.70110857  &  -2.37491751\\
C  &    1.01657867 &   2.07770634  &  -3.67060304\\
H  &    0.17335097 &   1.88929164  &  -4.32692528\\
H  &    1.28376973 &   3.13187528  &  -3.69594288\\
H  &    1.87451267 &   1.48688281  &  -3.98344684\\
H  &    2.41375208 &   1.56759346  &   1.51765978\\
N  &    0.23903094 &  -0.68283510  &   2.07377291\\
N  &    0.37695351 &  -2.85369992  &   2.52078342\\
C  &    0.60777974 &  -1.65629160  &   2.95874429\\
H  &    1.04258657 &  -1.43401742  &   3.92175961\\
C  &   -0.44758344 &   1.56573653  &   2.91281414\\
H  &   -0.54921705 &   2.60872746  &   2.61249804\\
H  &   -1.43398023 &   1.16972673  &   3.15486050\\
H  &    0.16610944 &   1.53500116  &   3.81595373\\
\hline
\end{tabular}
\end{table}

\begin{figure}[ht]
\label{fig-app:CAS_MO_planar_4pi}
\includegraphics[width=0.99\textwidth]{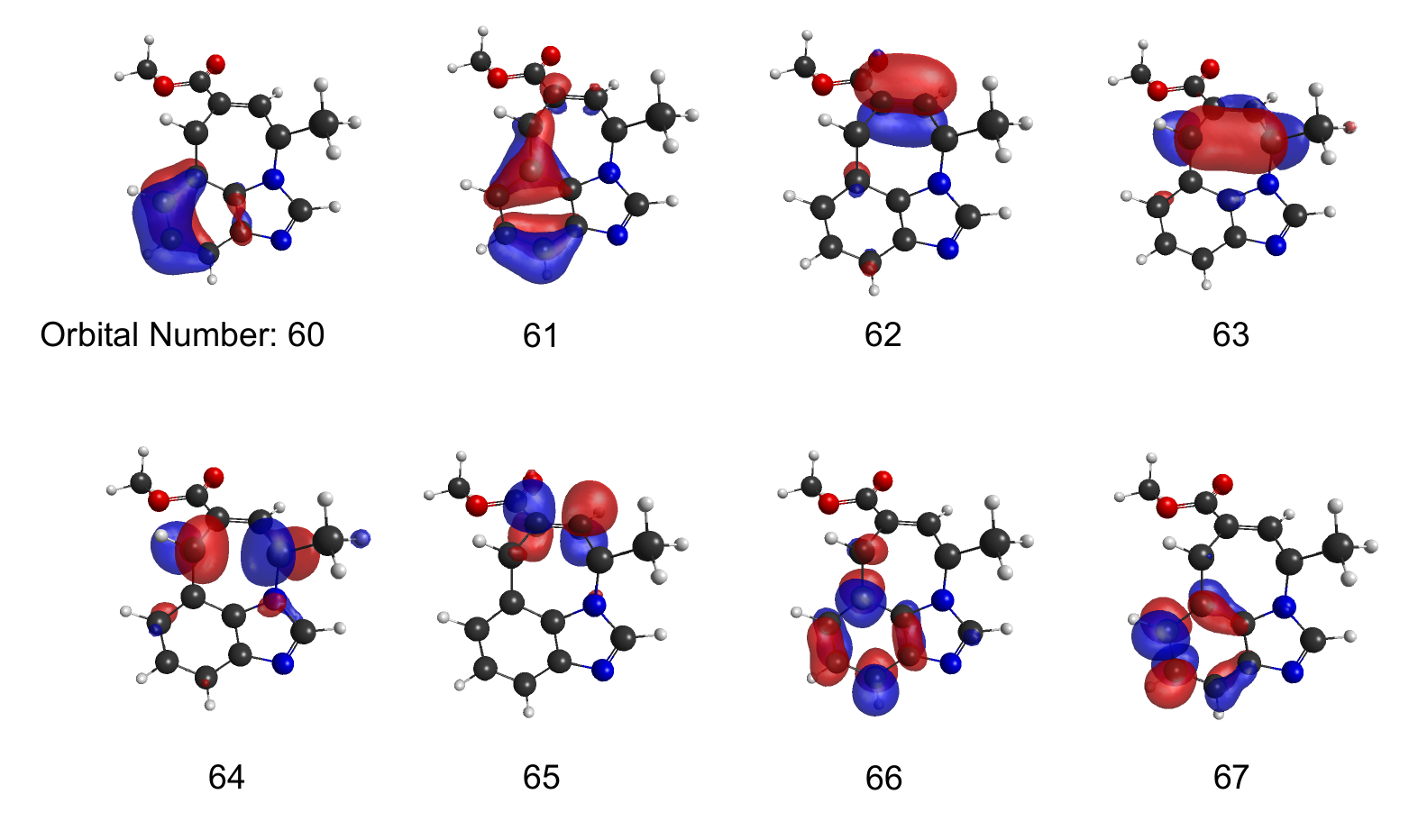}
\caption{The molecular orbitals in the active space (8,8) using def2-SVP basis sets in the optimized TS for the planar 4$\pi$ photoswitch reaction.}
\end{figure}

\end{document}